\documentclass[aps,amsmath,amssymb,nofootinbib,superscriptaddress,11pt]{revtex4}
\usepackage{graphicx,color}
\usepackage{epstopdf}

\begin{document}
\title{On the constituent counting rules for hard exclusive processes involving
multiquark states  }
\author{Feng-Kun Guo}
\email{fkguo@itp.ac.cn}
\affiliation{CAS Key Laboratory of Theoretical Physics, Institute of
Theoretical Physics, Chinese Academy of Sciences, Beijing 100190, China}

\author{Ulf-G. Mei\ss{}ner}
\email{meissner@hiskp.uni-bonn.de}
\affiliation{Helmholtz-Institut f\"ur Strahlen- und Kernphysik and
Bethe Center for Theoretical Physics, Universit\"at Bonn,  D--53115 Bonn,
Germany}
\affiliation{Institute for Advanced Simulation, Institut f{\"u}r
Kernphysik and J\"ulich Center for Hadron Physics, Forschungszentrum J{\"u}lich,
D-52425 J{\"u}lich, Germany}

\author{Wei Wang}
\email{wei.wang@sjtu.edu.cn}
\affiliation{
INPAC, Shanghai Key Laboratory for Particle Physics and Cosmology,
Department of Physics and Astronomy, Shanghai Jiao-Tong University,
Shanghai 200240,   China}

\begin{abstract}

At high energies, the cross section of  a hard
exclusive process at finite scattering angle falls off  as a negative power of the
center-of-mass energy $\sqrt{s}$.
If all involved quark-gluon  compositions undergo  hard momentum transfers, the
scaling of the fall-off is determined by the  underlying valence structures  of
the initial
and final states,  known as the constituent counting rules.
It was argued in the literature that the counting rules are a powerful tool  to
determine the valence degrees of freedom inside multiquark states when applied
to exclusive production processes.
However, we demonstrate that for hadrons with hidden flavors  the naive
application of the  constituent counting  rules is problematic, since  it is not
mandatory for all components  to participate the hard scattering at the scale
$\sqrt{s}$. The correct scaling rules can be obtained easily by using
effective field theory. A few examples involving the $Z_c(3900)^\pm$ and
$X(3872)$ are discussed.

\end{abstract}

\maketitle

\newpage

The concept of valance quarks has  played a major role in the classification  of
the hadronic states.
Hundreds of hadrons were discovered in experiments, and most  of them are
accommodated in the  quark model:  mesons and baryons are composed of a
quark--antiquark pair and three quarks, respectively. Here, the terminology
quark refers to the valence degrees of freedom. Hadrons beyond such
configurations are dubbed as exotic, and searching for them, in particular those
with exotic quantum numbers which cannot be formed by the above mentioned simple
configurations,  is of utmost importance in understanding the low-energy
nonperturbative quantum chromodynamics (QCD) because color confinement allows
such color-singlet states.\footnote{Classical large $N_c$ arguments state that
$q\bar q q\bar q$ tetraquark states are absent in the large $N_c$ limit (see
Refs.~\cite{Witten:1979kh,Coleman}). However, this conclusion was challenged in
Ref.~\cite{Weinberg:2013cfa} where it was argued that tetraquark states can
exist in the large $N_c$ limit with narrow widths. This was elaborated on in
Ref.~\cite{Knecht:2013yqa}.} Thanks to worldwide experiments   during the last
decade at  $e^+e^-$ and hadron colliders, a plethora of new structures as
candidates of various hadron resonances were reported with properties different
from quark model expectations, and it is probable that some of them could be
interpreted as exotic   multiquark states. Most of these new discoveries are in
the heavy quarkonium mass region (for reviews, see, e.g.,
Refs.~\cite{Brambilla:2010cs,Esposito:2014rxa,Chen:2016qju,Ali:2016gli}).
Among them, a milestone was the discovery of the $X(3872)$ by the Belle
Collaboration in 2003 and confirmed by several other experiments
later~\cite{Choi:2003ue,Acosta:2003zx,Aubert:2004ns,Abazov:2004kp}.
Since then one key topic in hadron physics is the study of these observed
structures.

Taking the $XYZ$ states in the charmonium mass region as an example, a number of
 interpretations have been proposed including normal quarkonia, hybrid states,
compact multiquark states, hadro-charmonia, hadron molecules and effects due to
kinematical singularities. Most of these interpretations are based on quark
model notations assuming explicitly or implicitly that the number of (valence)
quarks is well defined, even when discussing the production of multiquarks at
very high momentum transfer. Then, a central question discussing the observed
candidates of exotic hadrons (including proposing new measurements of the
properties of these states and searching for new structures) is:
how can one model-independently  determine the valence quark-gluon composition
of a hadron? In a special case of a hadron located very close to an $S$-wave
threshold of two other hadrons, one can in fact measure the valence hadron
component since hadrons being asymptotic states can go on shell.
However, it becomes very complicated when one tries to determine the quark-gluon
components. This is because of confinement which tells us that quarks and gluons
are not asymptotic states and can not be measured directly in experiments. Let
us define the numbers of quarks and antiquarks as $n_q$ and $n_{\bar q}$,
respectively. In a system with hidden flavor,  $n_q-n_{\bar q}$ is well defined
because of baryon number conservation, while $n_q+n_{\bar q}$ is not. In
particular, one does not expect the latter to take  a definite value for a given
hadron in processes happening at different energy scales. However, the latter is
the key quantity discussed in some papers in the literature, and we conclude
that the conclusions of these papers are thus model-dependent and sometimes
problematic. This argument can be made more clear by showing why the constituent
``counting rules'' fail for multiquark states in hard exclusive processes.

Recently, it has been argued  in
Refs.~\cite{Kawamura:2013iia,Kawamura:2013wfa,Blitz:2015nra,Brodsky:2015wza,
Chang:2015ioc} that  the differential cross section for high energy production
of multiquark states should scale with a certain power of $s$, the center-of-mass
energy squared, predicted based on their expected valence quark structures.
More explicitly,  for a generic process $a+b\to c+d$, the cross-section is
argued to obey  the
behavior~\cite{Kawamura:2013iia,Kawamura:2013wfa,Blitz:2015nra,Brodsky:2015wza,
Chang:2015ioc}:
\begin{eqnarray}
\frac{d\sigma}{dt} \sim s^{2-n} f(\theta_{cm}),  \label{eq:scaling}
\end{eqnarray}
with $n=n_{a}+n_b+n_c+n_d$. Here, $s$ and $t$ are the conventional Mandelstam
variables, $\theta_{cm}$ is the scattering angle in the center-of-mass frame,
and $n_h$ is the number of constituents in the particle $h$.
Here $a,b,c,d$ denote generic leptons, photons
or  hadrons including multiquarks. A fundamental particle like a quark, electron, or a photon has $n_i=1$.
An ordinary meson has $n_i=2$,  a meson-meson molecule or a compact tetraquark
has $n_i=4$, and a pentaquark has $n_i=5$, all of which  amount to  $n_q+n_{\bar
q}$ defined before.  The investigated   processes include $\pi^-+p\to K^0+
\Lambda(1405)$,  $\gamma+p\to
K^++\Lambda(1405)$~\cite{Kawamura:2013iia,Kawamura:2013wfa,Chang:2015ioc},
exclusive electron-positron annihilation~\cite{Blitz:2015nra,Brodsky:2015wza}
and so on.
Moreover, within the tetraquark framework,  the authors of
Ref.~\cite{Brodsky:2015wza} have argued that based on distinctive fall-offs in
$s$ of the cross sections,  it is possible to distinguish
whether the tetraquarks are segregated into di-meson molecules,
diquark-antidiquark pairs, or more democratically arranged four-quark states.

Were the constituent counting  rules  correct,   it would provide a very
powerful and straightforward  tool to access the valence  quark structures of
the exotic hadrons. But unfortunately as we have argued  above and will show in
more detail below,  for hadrons with hidden-flavor quarks it is problematic to
apply such a naive constituent counting rule.
To be explicit, we will first  consider a simpler example involving only
ordinary mesons, $e^+e^-\to VP$, with $V$ and $P$ denoting ordinary light flavor
vector and pseudoscalar mesons, respectively.
This reaction does not follow the naive scaling rule shown in
Eq.~(\ref{eq:scaling}). Then we will adopt the framework of effective field
theory  and point out the problems in the derivation of the misleading scaling
behavior in Eq.~(\ref{eq:scaling}).

At very high energy with $\sqrt s\gg \Lambda_{\rm QCD}$, exclusive processes can
be understood in QCD perturbation theory~\cite{Lepage:1980fj}.
The scaling behavior exists   in the factorization limit and can be formally
derived  by matching the full theory, QCD, to an effective field
theory.
When factorization is applicable, one can  formally separate the interactions
according to the involved scales:
\begin{eqnarray}
 T{\rm exp}\left[i\int d^4 x {\cal L}_\text{int}(x)\right] =  T{\rm exp}\left[i\int
 d^4 x {\cal L}_\text{int}(x)\right]_{>\mu}\times T{\rm exp}\left[i\int d^4 x {\cal
 L}_\text{int}(x)\right]_{<\mu}
\end{eqnarray}
where $T$ stands for time ordering, $\mu$ is the factorization scale and for
high-energy processes we should use  $\mu\sim\sqrt{s}$ in order to suppress the
large logarithms in   higher order  contributions.
Perturbation theory at high energies  allows one to express the matrix
elements of Heisenberg operators  in terms  of  free  local operators  and the
interaction terms.
By including the interaction, we have:
\begin{eqnarray}
 \langle f| {\cal O}_H(0)|i\rangle
 &=& \langle f|T\left[{\cal O} \times {\rm exp}[i\int d^4 x {\cal L}_\text{int}(x)]_{>\mu}\times T{\rm exp}[i\int d^4 x {\cal L}_\text{int}(x)]_{<\mu}\right]|i\rangle \nonumber\\
 &\sim&  \langle f|T\left[{\cal O}' \times  {\rm exp}[i\int d^4 x {\cal L}_\text{int}(x)]_{<\mu}\right]|i\rangle\nonumber\\
 &\equiv &  \langle f|{\cal O}'_{H,\mu}|i\rangle,
\end{eqnarray}
where in the last step we have formally  integrated out the interactions above
the factorization scale $\mu$ using the operator product expansion, and obtained a new set of generic low-energy
effective operators ${\cal O}'$.   The interaction below $\mu$ contains no information on the  $1/\sqrt{s}$ scaling and thus
the  scaling behavior  can be obtained by counting the pertinent  number of constituents   in the operator ${\cal O}'$.
It is possible  to include the  effects due to the
renormalization group and resummation of double logarithms known as   Sudakov
logarithms. For simplicity, we do not consider these effects here since the
leading  power behavior will be unaltered. An implication of the above analysis
is that the $s$-scaling of a process at high energies is given by that contained
in the operator $\mathcal{O}'$, and one gets the scaling easily by identifying
the number of lines attached to the vertex described by that operator, which is
depicted as a circled cross in the figures in the
examples to be discussed below, with the value of $n$ in Eq.~\eqref{eq:scaling}.

\begin{figure}[ht]
\begin{center}
\includegraphics[width=0.5\textwidth]{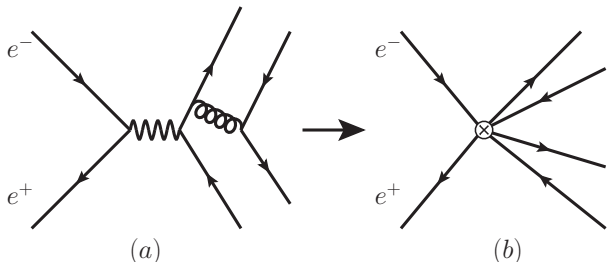}
\end{center}
\caption{Feynman diagrams for the $e^+e^-\to VP$ with the quark-antiquark pair produced by a   gluon.
(a) and (b) stand for the diagram in the full theory and the EFT, where the
hard propagators are shrank to a point depicted as a circled cross,
respectively.}
\label{fig:feynman1}
\end{figure}

It is straightforward to apply the constituent counting rules to the  $e^+e^-$
annihilation into two light mesons whose typical Feynman diagram is given in
Fig.~\ref{fig:feynman1}~(a). The $s$ power dependence is normally determined by
the constituent counting rule in
Eq.~(\ref{eq:scaling})~\cite{Lepage:1980fj} with $n=6$.
For the $e^+e^-\to VP$ with $V$ and $P$ being a vector and pseudoscalar
meson, respectively,
the differential  cross section scales as
\begin{eqnarray}
 \frac{d\sigma(e^+e^-\to VP)} {dt}  \propto \frac{1}{s^5}, \label{eq:eetoVPscaling}
\end{eqnarray}
which  differs from Eq.~(\ref{eq:scaling}) that  would give $1/s^4$ because of an
additional suppression factor of $1/s$ due to helicity flip.
Recent measurements of the process $e^+e^-\to KK^*$ by the Belle
Collaboration~\cite{Shen:2013okm} at 10.58~GeV  and CLEO
Collaboration~\cite{Adam:2004pr} are consistent with the above scaling in
Eq.~(\ref{eq:eetoVPscaling})  (see also results from BES~\cite{Ablikim:2004kv} and BaBar~\cite{Aubert:2006xw}).

\begin{figure}[th]
\begin{center}
\includegraphics[width=0.6\textwidth]{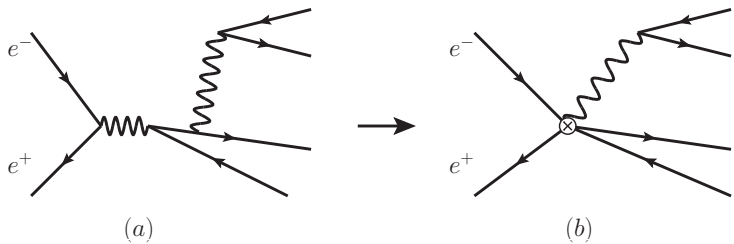}
\end{center}
\vskip -0cm
\caption{Feynman diagrams for the process $e^+e^-\to VP$ with both $V$ and $P$ being
neutral.  The neutral vector meson is produced via a photon.
Integrating out   high-off-shell propagators, one obtains (b) from (a).  }
\label{fig:feynman2}
\end{figure}

However, if  the vector meson is composed of a pair of quark and antiquark with
the same flavor, like the $\rho^0, \omega, \phi$ and $J/\psi$ mesons, the   scaling
behavior will be  different at  high energies. We show  a  production mechanism
in Fig.~\ref{fig:feynman2}~(a),
which  leads to the scaling behavior:
\begin{eqnarray}
 \frac{d\sigma(e^+e^-\to VP)} {dt}  \propto \frac{1}{s^3},
\end{eqnarray}
as can be read off from Fig.~\ref{fig:feynman2}~(b) where the effective
interaction vertex has $n=5$.
It is necessary to point out that this  production mechanism is suppressed by
the fine structure constant $\alpha_{\rm em}\sim 1/137$ and thus  less
important at low energies.
But apparently  at very high energies  this new diagram will provide the
dominant contribution and it gives a scaling rule different
from the one by naively counting the number of valence quarks in the
mesons~\cite{Lu:2007hr}.


It is   straightforward to understand
the above behaviors through the diagrams in Figs.~\ref{fig:feynman1} and
\ref{fig:feynman2}. In   Fig.~\ref{fig:feynman1}~(a), all internal
propagators have typically large off-shellness: $p^2\sim s$.   In
Fig.~\ref{fig:feynman2}~(a), the virtuality of the second photon, equal to the
mass square of the vector meson, is much smaller.
Thus, to accommodate with the constituent scaling  rule, one can technically
count the valence degrees of freedom of the neutral vector meson as $n_i=1$
since it is produced by a photon, which amounts to count the number of lines
attached to the effective vertex.
The lesson one can learn from the above example is: 
not all   ingredients undergo the hard momentum transfer at the scale
$\sqrt{s}$.
The scaling of the fall-off  is determined by the leading-power operator at the
scale $\mu=\sqrt{s}$ which has a nonzero matrix element with the hadron.
Actually, the original constituent counting rule is
applicable at finite scattering angles. If the scattering angle is small, at
least two of the involved particles are collinear which will also spoil the constituent counting rule.

\begin{figure}[th]
\begin{center}
\includegraphics[width=0.6\textwidth]{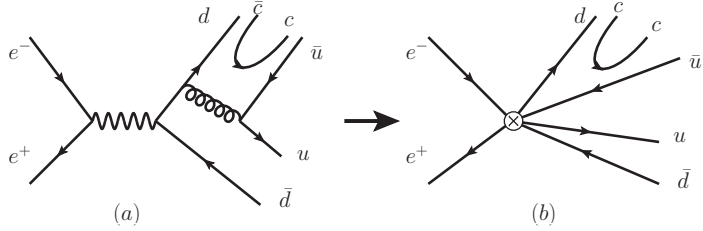}
\end{center}
\vskip -0cm
\caption{Feynman diagrams at leading power in $1/s$ for the reaction $e^+e^-\to
Z_c^-(\bar cc\bar ud) \pi^+$ at very high energies. The $\bar cc$ quark pair  is
generated by  QCD interactions at the  scale $\mu\sim m_c$ which is much lower
than $\sqrt{s}$.   }
\label{fig:Feynman2}
\end{figure}

Let us switch to the exclusive production of multiquark states, and take the reaction
$e^+e^-\to Z_c^\pm \pi^\mp$ as an example.  In Ref.~\cite{Brodsky:2015wza}, it
has been argued that its cross section  in the $s\to\infty$ limit obeys the scaling
\begin{eqnarray}
\frac{\sigma(e^+e^-\to Z_c^+(\bar cc\bar du)\pi^-(\bar ud))}  { \sigma(e^+e^-\to
\mu^+\mu^-)} \overset{?}{\propto} \frac{1}{s^4},
\end{eqnarray}
where the $Z_c^+(\bar cc\bar du)$ is a tetraquark state composed of two quarks
and two antiquarks.
We have put a
question mark to the above scaling behavior since we believe that it
is problematic at very high energies. We show a
production mechanism in Fig.~\ref{fig:Feynman2}~(a).  In this diagram,  the
heavy quark pair $\bar cc$ is generated from  the QCD vacuum, and thus such a
contribution  is suppressed by ${\cal O}(1/m_{c}^2)$. But since the main focus
of this work is the scaling behavior in terms of the collision energy,  we are
less interested in the $1/m_{c}^2$ suppression. Integrating out the off-shell
intermediate propagators at the scale $\sqrt{s}$ we find  that the $Z_c$
behaves as an ordinary $\bar qq$ meson and  the $s$ dependence scaling of the
cross-section is determined by the light quarks of the  $Z_c$:
\begin{eqnarray}
\frac{\sigma(e^+e^-\to Z_c^+\pi^-)} { \sigma(e^+e^-\to \mu^+\mu^-)} \propto
\frac{1}{s^2},
\end{eqnarray}
which can again be obtained by counting the number of lines attached to the
effective vertex. Apparently, this production mechanism will  become  dominant
at very high energies with $\sqrt s\gg m_c$.

A further example  is the reaction $e^+e^-\to Z_c^\pm Z_c^\mp$ argued to exhibit
the fall-off scaling~\cite{Brodsky:2015wza},
\begin{eqnarray}
\frac{\sigma(e^+e^-\to Z_c^+(\bar cc\bar du)Z_c^-(\bar cc\bar ud))}  {
\sigma(e^+e^-\to \mu^+\mu^-)} \overset{?}{\propto} \frac{1}{s^6},
\end{eqnarray}
which should be corrected as
\begin{eqnarray}
\frac{\sigma(e^+e^-\to Z_c^+(\bar cc\bar du)Z_c^-(\bar cc\bar ud))} { \sigma(e^+e^-\to \mu^+\mu^-)} \propto \frac{1}{s^2}.
\end{eqnarray}
The above results are applicable for the longitudinal polarization, while the
transverse polarized case should be further suppressed by $1/s$.

It is noteworthy to mention that  the above discussions on the $Z_c^\pm$
production are valid in  both the tetraquark (diquark-anti-diquark or a
democratically arranged four-quark state), and hadronic molecular pictures. The
scaling behavior is the same in both
scenarios, and therefore one can hardly make a statement distinguishing a
compact tetraquark from a meson-meson molecule using hard exclusive processes.

Ref.~\cite{Chang:2015ioc} has applied the naive counting rule to the
photoproduction of hyperon resonances  and attempted to study the
$\Lambda(1405)$. The $\Lambda(1405)$ is expected to be  a $\bar KN$-$\pi\Sigma$ bound
state~\cite{Dalitz:1959dn,Kaiser:1995eg,Oset:1997it,Oller:2000fj}.
 The fitted constituent number is energy-dependent, which can be
understood since the constituent counting rule determines the asymptotic behavior in the
large energy limit, and will be distorted by  finite energy corrections.  At the
largest collision energy, however the obtained constituent number is consistent
with three for the $\Lambda(1405)$, despite of the large errors.  The fact  that $n=3$
does not imply that the $\Lambda(1405)$ is an ordinary $uds$ baryon but instead
it shows that  the production mechanism of the  $\Lambda(1405)$  at short
distances involves  three quarks.

Regarding the notable exotics candidate $X(3872)$,
an important task  in understanding its  nature  involves the
discrimination of a quark-antiquark configuration,  a compact multiquark
configuration and a hadronic molecule (we refer to the recent
review~\cite{Chen:2016qju} which summarizes nicely the literature).
Unlike the $Z_c^\pm$, the $X(3872)$ is neutral, and both
the light quark--antiquark pair and charm--anticharm quark pair are hidden.  So
in hard exclusive processes, the $X(3872)$ can be produced at short distances by
two sets of operators:
\begin{eqnarray}
 \langle X|\bar c\Gamma c|0\rangle, \;\;\;
 \langle X|\bar q\Gamma q|0\rangle,
\end{eqnarray}
with $q$ being a light  $u/d$ quark field, and $\Gamma$ denoting the Lorentz
structure of the operator to produce the $X$.
The explicit form of the  contributing operators  depends on the process.
For instance,  Ref.~\cite{Braaten:2004jg} has  explored  the inclusive production of $X(3872)$  in $B$ decays and
at hadron colliders, and pointed out that  the most important term in the factorization formula should be the color-octet $^3S_1$ term.
In   exclusive $B_c$ decays into   $X(3872)$, the  $ \langle
X|\bar c\Gamma c|0\rangle$ contributes~\cite{Wang:2015rcz}, and ratios of
branching fractions can be  predicted with a high precision under this
mechanism, no matter whether the long-distance nature of    $X$ is given by   a tetraquark or
hadronic molecule composition.

To understand the internal structure of  exotic hadron candidates, it is
essential to figure out the corresponding  valence quark-gluon compositions.
However, at low energies, since the effective degrees of freedom are  hadrons,
and only integrated quantities can be observed,  it is very hard to determine
the valence components.
One might hope that  cross sections of the exclusive
productions of these hadrons may be used to determine the valence components of
a multiquark state since there are constituent counting rules for hard
exclusive processes determined by the number of elementary particles involved.
However, we have argued in this paper that multiquark states with  hidden
flavors do not have to follow the scaling rule by  simply counting the number of valence
quarks and antiquarks. The reason is that the hidden-flavor pair could
be produced by a much softer momentum exchange. In the spirit of effective field
theory, the correct counting rule can be obtained by integrating out the hard
scale, modulo possible additional factors due to helicity suppression and so
on. The counting rule is demonstrated using the  $e^+e^-\to VP$ process. We
discussed   productions of the $Z_c^\pm(\bar
cc\bar ud/\bar cc\bar du)$, $X(3872)$ and others  in a few hard exclusive
processes.

\section*{Acknowledgements}

WW would like to acknowledge  Z.-G.~He, R.~F.~Lebed, H.-N.~Li, C.-D.~L\"u, Y.-Q.~Ma, Y.~Jia,
D.-S.~Yang, and  Q.~Zhao for valuable discussions especially  during the
workshop ``QCD study group 2016" at Shanghai JiaoTong University.  The authors
have benefited a lot from  the program  ``Clustering effects of nucleons in
nuclei and quarks in multi-quark states" at the Kavli Institute for Theoretical
Physics China at the Chinese Academy of Sciences.  FKG and WW gratefully
acknowledge the hospitality at the HISKP where part of this work was done. This
work is supported  in part by DFG and NSFC through funds provided to the
Sino-German CRC 110 ``Symmetries and the Emergence of Structure in QCD" (NSFC
Grant No.
11261130311). FKG and WW are also supported by the Thousand Talents Plan for
Young Professionals. The work of UGM was supported in part by The Chinese
Academy of Sciences (CAS) President's International Fellowship Initiative (PIFI)
with grant no. 2015VMA076.  WW is supported in part  by National  Natural
Science Foundation of China under Grant
 No.11575110,  Natural  Science Foundation of Shanghai under Grant  No.
15DZ2272100 and No. 15ZR1423100,  by the Open Project Program of State Key
Laboratory of Theoretical Physics, Institute of Theoretical Physics, Chinese
Academy of Sciences, China (No.Y5KF111CJ1), and  by   Scientific Research
Foundation for   Returned Overseas Chinese Scholars, State Education Ministry.

\end{document}